\documentclass[useAMS,usenatbib]{mn2e}

\usepackage{natbib, aas_macros}
\usepackage[dvips]{graphicx}
\usepackage{wrapfig}
\graphicspath{{./figs/}}
\usepackage{color}
\usepackage{amsmath}
\usepackage{amssymb}

\newcommand{\Msun}{M_{\odot}}
\newcommand{\Mtot}{M_{\rm tot}}
\newcommand{\Zsun}{Z_{\odot}}

\newcommand{\A}{\rm \AA}

\newcommand{\EBV}{E{(B-V)}}
\newcommand{\meanEBV}{\left< E{(B-V)} \right>}
\newcommand{\Mab}{M_{1700}}
\newcommand{\Qabs}{Q_{{\rm abs}, \lambda}}
\newcommand{\rd}{r_{\rm d}}
\newcommand{\tauuv}{\tau_{1700}}

\newcommand{\fdd}{f_{\rm d}}

\title[Dust properties of Lyman break galaxies in cosmological simulations]
{Dust properties of Lyman break galaxies in cosmological simulations}
\author[Yajima et al.]
{Hidenobu Yajima$^{1, 2, 3}$\thanks{E-mail: yajima@roe.ac.uk (HY)}, Kentaro Nagamine$^{4, 5}$, Robert Thompson$^{4, 6}$, Jun-Hwan Choi$^{7, 8}$
\\
$^{1}$ SUPA\thanks{Scottish Universities Physics Alliance}, Institute for Astronomy, University of Edinburgh, Royal Observatory, Edinburgh, EH9 3HJ, UK\\
$^{2}$Department of Astronomy and Astrophysics, Pennsylvania State University,
525 Davey Lab, University Park, PA 16802, U.S.A.\\
$^{3}$Institute for Gravitation and the Cosmos, The Pennsylvania State University, University Park, PA 16802, U.S.A.\\
$^{4}$Department of Physics and Astronomy, University of Nevada Las Vegas, 4505 S. Maryland Pkwy, Las Vegas, NV 89154-4002, U.S.A.\\
$^{5}$Department of Earth and Space Science, Graduate School of Science, Osaka University, 1-1 Machikaneyama-cho, Toyonaka, Osaka, \\
560-0043, Japan\\
$^{6}$Department of Astronomy, University of Arizona, 933 N. Cherry Ave., Tuscon, AZ 85721-0064, U.S.A.\\
$^7$Department of Physics \& Astronomy, University of Kentucky, Lexington, KY 40506-0055, U.S.A.\\
$^8$Department of Astronomy, University of Texas, Austin, TX 78712-1205, U.S.A.
}
 
\begin{document}

\date{Accepted ?; Received ??; in original form ???}

\pagerange{\pageref{firstpage}--\pageref{lastpage}} \pubyear{2008}

\maketitle

\label{firstpage}

%
%
\begin{abstract}
Recent observations have indicated the existence of dust in high-redshift galaxies, however, the dust properties in them are still unknown. 
Here we present theoretical constraints on dust properties in Lyman break galaxies (LBGs) at $z=3$
by post-processing a cosmological smoothed particle hydrodynamics simulation with radiative transfer calculations. 
We calculate the dust extinction in 2800 dark matter halos using the metallicity information of individual gas particles in our simulation. 
We use only bright galaxies with rest-frame UV magnitude $M_{\rm 1700} < -20$\,mag, and study the dust size, dust-to-metal mass ratio, and dust composition. 
From the comparison of calculated color excess between $B$ and $V$-band (i.e., $\EBV$) and the observations, we constrain the typical dust size, and show that the best-fitting dust grain size is $\sim 0.05\, \mu$m, which is consistent with the results of theoretical dust models for Type-II supernova.  
Our simulation with the dust extinction effect can naturally reproduce the observed rest-frame UV luminosity function of LBGs at $z=3$ without assuming an ad hoc constant extinction value. 
In addition, in order to reproduce the observed mean $\EBV$, we find that the dust-to-metal mass ratio 
needs to be similar to that of the local galaxies, and that the graphite dust is dominant or at least occupy half of dust mass. 
\end{abstract}

%
%

%
%
\section{Introduction}
Recent observations by large, optical telescopes have detected numerous high-redshift 
galaxies utilizing the drop-out technique at the Lyman-limit wavelength, 
and these galaxies are called the Lyman break galaxies (LBGs) \citep[e.g.,][]{Steidel96, Steidel03, Shapley03, Adelberger04, Ouchi04}. 
The spectral energy distributions (SEDs) of these LBGs indicate that the ultra-violet (UV) fluxes are reduced due to dust extinction. 
Because the cross section of dust varies as a function of wavelength, 
the dust extinction modifies the color of galaxies from the intrinsic one \citep{Calzetti00}. 
Recent LBG surveys presented the color excess of $B$ and $V$ band, $\EBV$,
and showed that some fraction of UV continuum flux might be absorbed by interstellar dust \citep[e.g.,][]{Ouchi04, Iwata07, Bouwens07, Reddy08, Reddy09, Bouwens10, Shapley11}. 

At high redshift, Type-II supernovae (SNe) are considered to be the main contributor of dust production, because other production mechanisms, such as the stellar wind from AGB stars and condensation in molecular clouds, are inefficient \citep{Bianchi07, Zhukovska08}. 
The production rate of dust is thus related to the formation rate of massive stars, and also indirectly to the initial mass function (IMF) and star formation rate (SFR). 
Dust can also assist star formation through efficient formation of hydrogen molecules on its surface \citep[e.g.,][]{Omukai08, Schneider10, Dopcke11, Yamasawa11}. 
On the other hand, Type-II SNe can destroy dust by sputtering effect \citep{Schneider04, Nozawa06, Nozawa07},  
where the high-temperature gas particles collide with dust and strip off components from dust.  
In addition, dust-dust collision in magnetic fields, i.e., the `shattering effect' also destroys dust \citep{Hirashita10, Hirashita13}.
The destruction efficiencies by these processes sensitively depend on the physical conditions in the interstellar medium (ISM), 
e.g., local temperature, density, and magnetic field strength. 
Interstellar dust is a key component in understanding the star formation history, physical properties of galaxies, and stellar IMF. 
Nevertheless the dust properties in high-redshift galaxies are still not well understood. 

In the examination of dust effects on star formation and extinction of stellar radiation, 
the typical dust size is the key factor because it determines the total dust surface area. 
The dust size distribution will be determined by the combination of different physical processes such as the formation, sputtering \& shattering effect, and it is expected to evolve as a function of time and space. 

Inside the shock wave from SNe, only small dust grains with $\rd \lesssim 0.01 ~\micron$ 
($\rd$ is the radius of dust) 
are captured by the shock and destroyed due to the sputtering process \citep{Nozawa07}.
As a result, the initial dust size is likely to be $\rd \gtrsim 0.01~\micron$,
and the typical dust size may be $\rd \sim 0.05~\micron$ \citep{Todini01, Dayal11}.
After the initial production of dust by Type-II SNe, 
very small dust ($\rd < 0.01~\micron$) can be produced by the shattering effect, and the dust size distribution would shift to a steeper power-law index of $\alpha \sim -3.5$ from the initial slope of $\alpha \sim -3.0$ \citep{Nozawa07, Hirashita13},  
where 
$ \frac{dn_{\rm d}}{d\rd} \propto \rd^{\alpha}, $
and $n_{\rm d}$ is the number density of dust.
This power-law slope agrees with the observations of local galaxies.

Unfortunately it is too difficult to follow the dust formation and destruction processes in galaxies completely even with today's state-of-the-art computational facilities and calculation codes, because it requires the following of multiple massive star formation, shock propagations of Type-II SNe, magnetic field and turbulence in ISM over a wide spatial range from sub-pc to kpc.
Due to this difficulty, previous works have mostly adopted simple models of dust assuming the same properties as the local galaxies or single dust size to reproduce radiation properties of observed high-redshift galaxies.
For example, \citet{Nagamine04e, Night06, Nagamine10b}  and \citet{Jaacks12} reproduced the rest-frame UV luminosity functions (LFs) of LBGs at $z\ge 3$ by cosmological SPH simulations with a simple extinction model that assumes the extinction law of local starburst galaxies \citep{Calzetti00} and a constant color excess. 
\citet{Dayal09} used a slab dust model with the graphite dust size of $0.05~\micron$, and confirmed that the simulated galaxies at $z=5.7$ agreed with the observed $\EBV$.
In addition, \citet{Dayal11} simulated LFs of Lyman-$\alpha$ emitters (LAEs) at $z=5.7$ by combining cosmological SPH simulations with Monte Carlo radiative transfer (RT), and showed that the dust attenuation with clumpy distribution was needed to reproduce the observational properties. 
\citet{Kobayashi10} reproduced the radiation properties of LAEs at $z=3.1-5.7$ using a semi-analytical model and a slab dust model with the Milky Way (MW) dust extinction curve.
\citet{Gonzalez-Perez13} investigated the dust attenuation of LBGs in a semi-analytical model using a simple radiative transfer model. 
They assumed the bulge and disk follow the MW and the Small Magellan Could (SMC) dust model, 
and showed that brighter LBGs suffered stronger dust attenuation than the fainter ones.
Very recently \citet{Kimm13} carried out Monte Carlo RT calculations for $z=7$ simulated galaxies in cosmological AMR simulation, and reproduced the observed UV spectral slope $\beta$.  They concluded that the dust model of SMC was favored for the high-redshift galaxies rather than the MW dust model. 

In the present work, we attempt to constrain the typical dust size by comparing the results of RT calculations with  observational data. 
Depending on the dust properties, the value of $\EBV$ and galaxy LF will change. 
The observations have measured $\EBV$ of LBGs, and showed that the typical value is $\EBV \sim 0.14$ at $z \sim 3$ \citep[e.g.,][]{Shapley01, Reddy08}.  

Our paper is organized as follows.
We describe our simulations and our approach for RT calculations in Section~\ref{sec:model}. 
In Section~\ref{sec:result}, we present our results, and show simulated LF with extinction, probability distribution function of $\EBV$
and the dependence of $\EBV$ on the mass and UV flux of galaxies. 
In Section~\ref{sec:discussion}, we investigate the dust-to-metal mass ratio and the dust composition. 
Finally, we summarize our main conclusions in Section~\ref{sec:summary}.
We focus on redshift $z=3$ in this paper.

%
%
\section{Simulation, Dust Model, \& Radiative Transfer Method}
\label{sec:model}

We use a modified version of the smoothed particle hydrodynamics (SPH) code GADGET-3 \citep[originally described in ][]{Springel05e}.
The details of the simulation were presented in \citet{Thompson13}.
Here, we briefly summarize the simulation setup. 
Our code includes the heating by a uniform UV background \citep{Faucher09}, 
cooling by hydrogen, helium and metals \citep{Choi09},  
the self-shielding effect of the UV background radiation \citep{Nagamine10a, Yajima12d}, and 
the "Multicomponent Variable Velocity (MVV)" wind model \citep{Choi11}, which is the hybrid of energy-driven and momentum-driven wind models.  
In addition, the star formation (SF) model in our simulations is based on the H$_{2}$ mass in each SPH particle, which has many advantages over the previous SF models  \citep{Thompson13}.

The calculations are done with total $2 \times 400^{3}$ particles for both gas and dark matter, 
with a box size of comoving $34 h^{-1}$\,Mpc. 
The particle masses are $m_{\rm DM}=3.6\times 10^{7}~h^{-1} \Msun$ and $m_{\rm gas}=7.3\times10^{6}~h^{-1} \Msun$.
The comoving gravitational softening length is $\epsilon = 3.4 ~h^{-1}$\,kpc. 
We adopt the cosmological parameters that are consistent with the best-fit of WMAP7 data: 
$\Omega_{\rm m}=0.26, \Omega_{\Lambda}=0.74, \Omega_{\rm b}=0.044,$ $h=0.72, \sigma_{8}=0.80, n_{\rm s}=0.96$ \citep{Komatsu11}.

After the cosmological hydrodynamic simulation is finished, we carry out the RT calculations as post-processing. 
First, we place a uniform grid around each dark matter halo with a size of the virial radius. 
The size of each grid cell is the same as the comoving gravitational softening length $\epsilon$. 
The gas and metal density of each cell are estimated from neighboring SPH particles with the kernel function (i.e. SPH smoothing). 
Then we calculate the radiative transfer (ray tracing) of stellar radiation in the dusty ISM, where the dust mass is assumed to be proportional to the metal mass.  
In this work, we shoot only one ray towards one particular viewing angle from each star particle in order to save computational time, i.e., not the angular mean flux. 
A similar post-processing RT method was also used in \citet{Yajima09, Yajima11, Yajima12a, Yajima12d}. 

From the multi-band observations of local galaxies, 
\citet{Draine07} showed that the dust mass can be estimated from gas mass and metallicity with the constant dust-to-metal mass ratio: 
\begin{equation}
M_{\rm dust} \sim \kappa M_{\rm gas} \left( \frac{Z}{\Zsun} \right ), 
\label{eq:dustgas}
\end{equation}
where $\kappa \approx 0.01$ is the dust-to-gas mass ratio at solar metallicity, $Z$ is the metallicity, and $\Zsun$ is the solar metallicity. 
We determine the dust mass in each grid cell using the above $\kappa$ and the simulated metal mass in each cell.  In our calculation, the metal-to-gas mass ratio is taken to be 0.02 at solar metallicity, 
hence the dust-to-metal mass ratio of our fiducial model is $M_{\rm dust} / M_{\rm metal} =  0.5$. 
Different dust-to-metal mass ratios will be examined in Section~\ref{sec:discussion}.

\begin{figure}
\begin{center}
\includegraphics[scale=0.35]{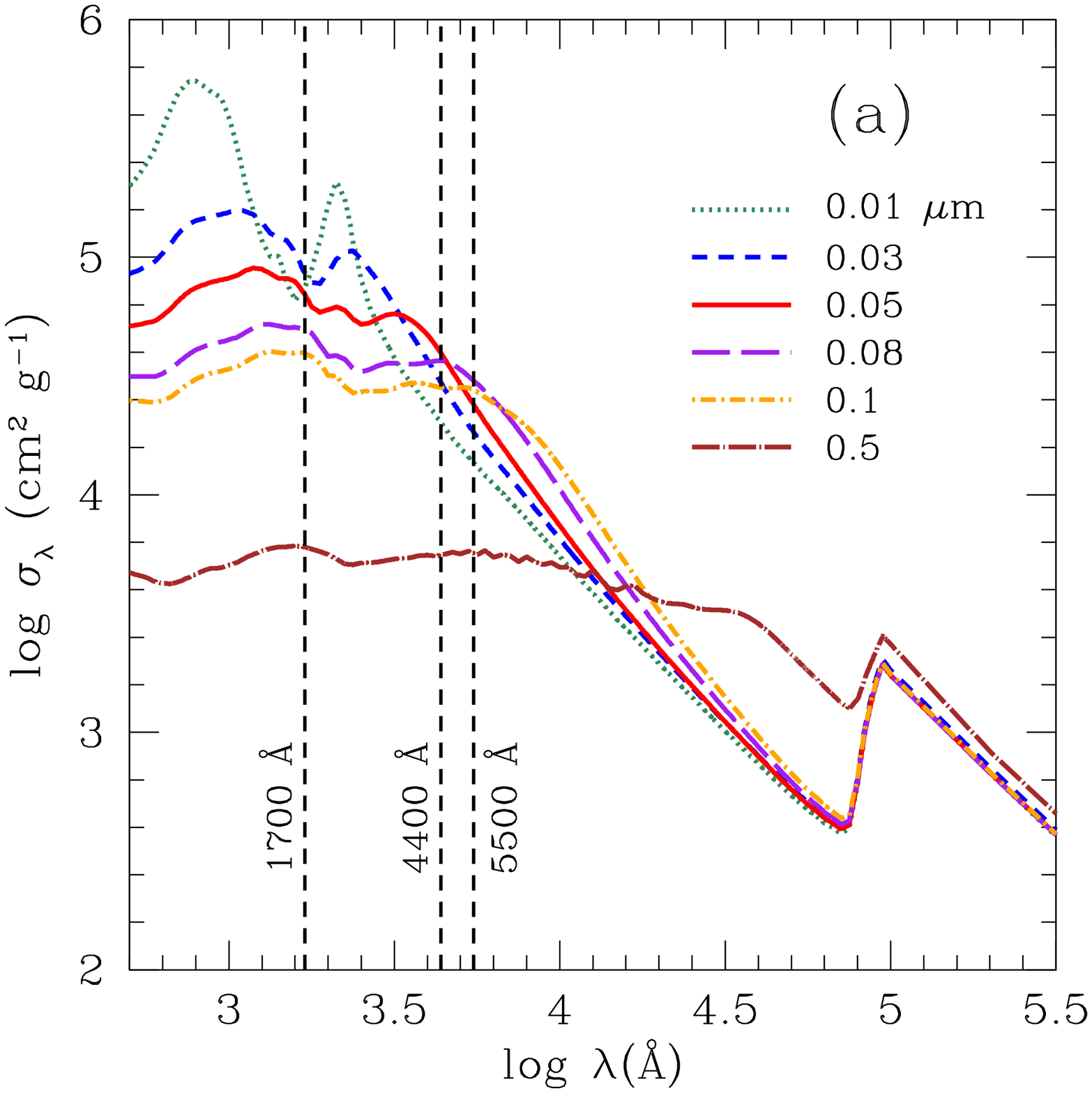}
\includegraphics[scale=0.35]{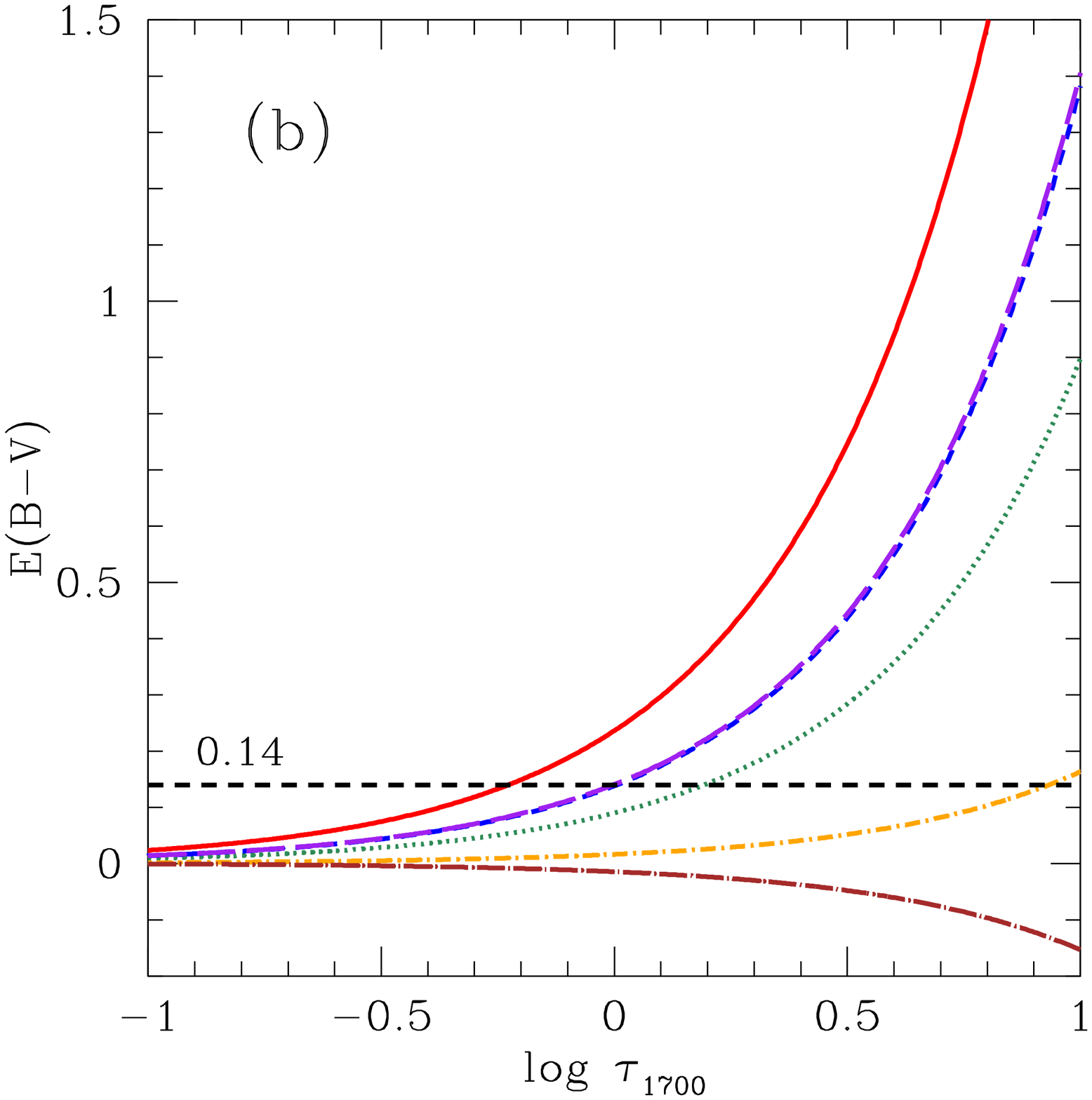}
\caption{
{\it Panel (a):} Absorption efficiency per unit dust mass $\sigma_\lambda$ as a function of wavelength. 
Different colors and line types represent different radius of dust.  
The spherical dust is assumed as described in the text. 
{\it Panel (b):}  Color excess $E(B-V)$ as a function of optical depth at $1700~\rm \A$ ($\tau_{1700})$. 
}
\label{fig:cross}
\end{center}
\end{figure}

The dust extinction curve also depends on the dust composition. 
It is believed that graphite and silicate are the main components of dust in galaxies, based on the observations of $9.7~\micron$ absorption feature of silicate and $2175~\rm \AA$ of graphite \citep{Laor93}.
In this work, we assume that the dust consists of graphite and silicate with $1:1$ mass ratio.

We calculate the multi-wavelength radiative transfer along the ray via 
\begin{equation}
I_{\nu} = I_{\nu}^{0}{e}^{-\tau_{\lambda}}.
\end{equation}
The ray is shot from all star particles in the grid to the grid boundary, corresponding to the size of the virial radius of each halo. 
We compute the intrinsic spectral energy distribution (SED) of each star particle with the \citet{Salpeter55} IMF using the population synthesis code P\'{E}GASE v2.0 \citep{Fioc97}, based on the mass, formation time and metallicity of each star particle. 
The Salpeter IMF was commonly used in the analysis of observed LBGs \citep[e.g.,][]{Ouchi04, Reddy08},
while some works and our original hydrodynamics simulations assumed the Chabrier IMF \citep{Chabrier03}. 
However, since we mainly focus on the color excess, our results are not very sensitive to the IMF. 
On the other hand, rest frame UV flux can change between Salpeter and Chabrier IMF by a factor $\sim 1.58$ \citep{Treyer07}.  
Hence, our analysis of UV LF can shift to the brighter side by $\sim 0.5$ mag if we switch to the Chabrier IMF,  but it is still in the error bars of the observations and does not change our main conclusions. 

The optical depth is estimated by
\begin{equation}
d\tau_{\lambda} = (\Qabs^{\rm s} \pi \rd^{2} n_{\rm d}^{\rm s} 
+ \Qabs^{\rm g} \pi \rd^{2} n_{\rm d}^{\rm g}) ds,
\end{equation}
where $\Qabs$ is the dust absorption coefficient, $\rd$ and $n_{\rm d}$ are the dust radius and number density, respectively.
Here we assume that the dust particles are spherical, hence the optical depth becomes $\tau_{\nu} \propto \Qabs / \rd$. 
For a given amount of dust mass, the dust number density increases with decreasing size. 
We use the $\Qabs$ data of astronomical silicate and graphite dust in \citet{Draine84} and  \citet{Laor93}.

\begin{figure*}
\begin{center}
\includegraphics[scale=0.75]{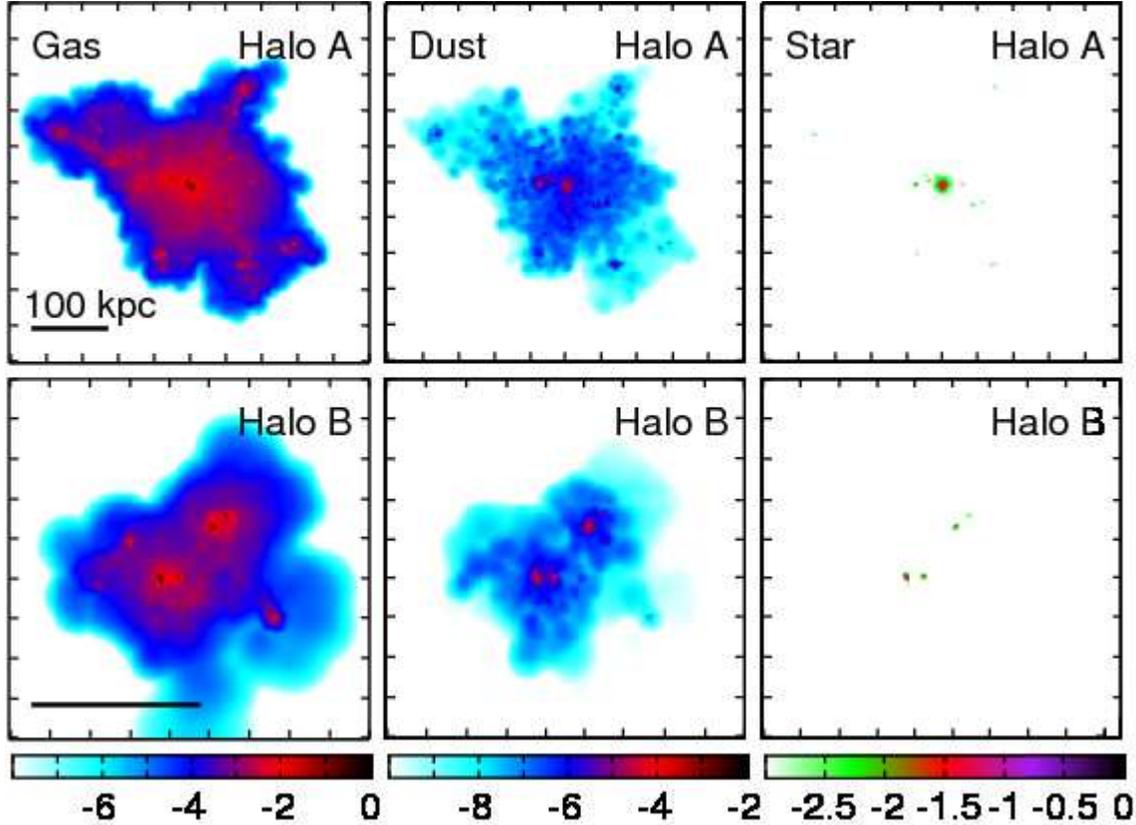}
\caption{
Projected density maps of gas ({\it left}), dust ({\it middle}) and stars ({\it right}) for the two representative, massive halos in our cosmological SPH simulation at $z=3$. 
The color gradient shows the projected density in log scale in units of [g\,cm$^{-2}$].
The total masses of halos A and B are $M_{\rm tot} = 2.2 \times 10^{12}$ and $5.1 \times 10^{11}~\rm\Msun$, respectively, which are appropriate for typical Lyman-break galaxies.
The virial radii are $100.8$\,kpc (halo A) and $61.9$\,kpc (halo B) respectively. 
The scale of physical 100\,kpc is shown by the black bar.  
}
\label{fig:img}
\end{center}
\end{figure*}

We also define the dust absorption efficiency per unit mass $(\sigma_{\lambda})$ via 
\begin{equation}
d\tau_{\lambda} = \sigma_\lambda\, m_{\rm d}\, ds,
\end{equation}
where $m_{\rm d}$ is the dust mass per unit volume. 
In the spherical dust model, 
\begin{equation}
m_{\rm d} = \frac{4}{3}\pi \rd^3\, \rho\, n_{\rm d},
\end{equation}
where $\rho$ is the mass density in each dust grain.  Combining these relationships, we find that $\sigma_\lambda$ is proportional to $\Qabs$ via 
\begin{equation}
\sigma_\lambda = \frac{3}{4} (\rho \rd)^{-1} \Qabs. 
\label{eq:sigma}
\end{equation}

The values of $\sigma_\lambda$ and $\Qabs$ depend on the wavelength sensitively, 
and the wavelength dependency of $\sigma_\lambda$ is shown in Figure~\ref{fig:cross}(a).
For $\lambda \lesssim 2\pi \rd$, the values of $\Qabs$ is nearly constant and unity,
while it decreases with increasing wavelength by $\lambda^{-1}$ at $\lambda > 2\pi \rd$ (likewise for $\sigma_\lambda$).
From Equation~(\ref{eq:sigma}), one can see that $\sigma_\lambda$ of smaller dust  is higher than that of the larger dust at shorter wavelengths.
Figure~\ref{fig:cross}(a) also shows that $\sigma_\lambda$ becomes degenerate at longer wavelength for all dust sizes, except for the very large dust size of $\rd = 0.5 \micron$. 

Due to the difference in $\sigma_\lambda$ at different wavelengths,
the color excess between $B$ and $V$-band, $\EBV$, arises. 
Figure~\ref{fig:cross}(b) shows $\EBV$ as a function of optical depth at $1700~\A$ ($\tauuv$). 
We use $4400\,\rm\AA$ and $5500\,\rm\AA$ as the representative of $B$ and $V$-band, and 
the value of $\EBV$ is estimated by 
\begin{equation}
\EBV = - 2.5 \left[ \log \left( \frac{F_{4400}^{\rm RT}}{F_{4400}^{\rm Int}} \right) -  \log \left( \frac{F_{5500}^{\rm RT}}{F_{5500}^{\rm Int}} \right) 
\right], 
\label{eq:ebv}
\end{equation}
where $F_{4400}^{\rm Int}$ and $F_{5500}^{\rm Int}$ are flux densities ($\rm erg~s^{-1}~cm^{-2}~Hz^{-1}$) at $4400$ and $5500~\A$ without dust extinction respectively, 
and $F_{4400}^{\rm RT}$ and $F_{5500}^{\rm RT}$ are the ones with dust extinction. 
It is difficult to achieve high $\EBV$ with large dust sizes of $\gtrsim 0.1~\micron$. 
Since $4400~\rm \A$ and $5500~\rm \A$ are both shorter than $2 \pi \rd$, 
the values of $\sigma_\lambda$ are very close to each other at these two wavelengths for dust size of $\gtrsim 0.1\,\micron$.
From Figure~\ref{fig:cross}(b), we see that we need $\tauuv \sim 10$ to achieve $\EBV \sim 0.14$ with $0.1\,\micron$ dust.
In the case of $0.5~\micron$ dust, the value of $\sigma_\lambda$ and $\Qabs$ are almost constant up to $\sim 3\,\micron$, therefore it results in negligible $\EBV$ even at quite high dust column density.
For $0.05~\micron$ dust, the absorption efficiency is somewhat higher than that of smaller dust at $\sim 3000~\A$, 
and then it starts to decrease with increasing wavelength. 
This causes somewhat higher $\EBV$ for $0.05~\micron$ dust, even though 
the slope of absorption efficiency at $\lambda > 2 \pi \rd$ does not depend on the size with $\sigma \propto \lambda^{-1}$.  


%
%

\section{Results}
\label{sec:result}


\begin{figure*}
\begin{center}
\includegraphics[scale=0.7]{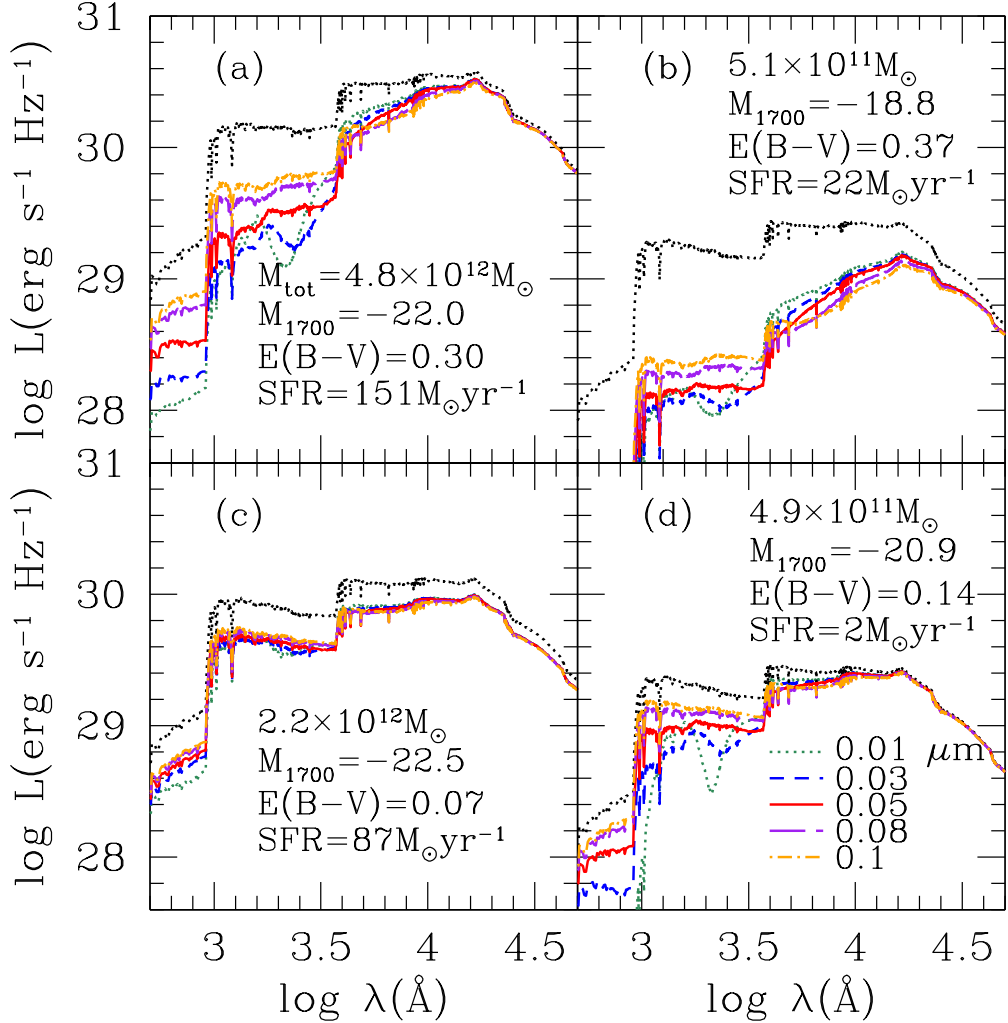}
\caption{
The SEDs of four simulated galaxies, processed with different dust size models. 
The SEDs shown in panels ($a$) \& ($b$) correspond to the same halos as Halo A \& B shown in Figure~\ref{fig:img}.  The dotted lines are the SEDs without dust extinction. 
The colored lines indicate different dust size models. 
The values of $\Mab$ and $\EBV$ are estimated for the dust model of $\rd = 0.05~\rm \mu m$. 
}
\label{fig:sed}
\end{center}
\end{figure*}

\subsection{Distribution of Dust and Stars}
The projected distribution of dust and stars are shown in Figure~\ref{fig:img} for the two representative
massive halos at $z=3$ in our cosmological SPH simulation. 
The total masses of Halos A and B are $M_{\rm tot} = 2.2 \times 10^{12}$ and $5.1 \times 10^{11}~\rm\Msun$, respectively, 
which are appropriate for typical Lyman-break galaxies (LBGs).
Each of these two halos contain a few LBGs as shown by the stellar concentrations in the right panels. 
The gas distribution around stars is extended, and Halo A shows a filamentary-like gas structure, which resembles the so-called cold flow. On the other hand,
the dust distribution is strongly concentrated near the stellar component as shown by the red clumps in the middle panels. 
However, at the same time, the diffuse dust is also distributed widely in the halo as shown by the blue clouds.  
These dust and metals are ejected from the galaxies by galactic wind feedback up to $\sim 100$\,kpc, and the validity of our hybrid MVV wind model was examined by \citet{Choi11} via galactic wind speeds and C{\sc iv} statistics in the intergalactic medium.  
These LBGs in the same halo are in the process of merging, with relatively high total star formation rates of  
$151~ \Msun {\rm yr}^{-1}$ (Halo A) and $22~ \Msun {\rm yr}^{-1}$ (Halo B).
The total stellar masses in each halo is $1.4 \times10^{11} ~\Msun$ (Halo A) and  $1.7 \times10^{10} ~\Msun$ (Halo B),  and the average metallicity of all star particles in each halo is about $0.11 ~Z/\Zsun$. 
From these values, we confirm that our galaxy sample satisfies the physical properties of observed LBGs, e.g., star formation rate, metallicity, and halo mass \citep[e.g.,][]{Pettini01, Shapley01, Adelberger05, Mannucci09}.


\subsection{SEDs of four example halos \& optical depth distribution}

We examine the spectral energy distribution (SED) of four examples including the halos in Figure~\ref{fig:img} with dust extinction. 
The SED of a galaxy shows how much energy is coming out at each wavelength, and comparing its shape before and after the dust extinction highlights the absorption of UV photons by dust.  

Figure~\ref{fig:sed} compares the SEDs of four simulated galaxies before and after the dust extinction effect for different dust sizes. 
Some flux at UV to optical wavelengths are absorbed by dust. 
Of course, the dust extinction becomes smaller at longer wavelength
due to smaller cross section, and the dust extinction becomes negligible at $\lambda \gtrsim 1~\micron$. 
The emergent SEDs change with dust size. 
In particular, for $0.01\,\micron$ dust, there are dips in the UV range 
due to the bump in $\sigma_{\lambda}$ shown in Figure~\ref{fig:cross}.

The dust extinction is significantly different between the galaxies, because of the complicated distribution of dust and stars as shown in Figure~\ref{fig:img}.
The high column density gas absorbs stellar radiation over a wide range of wavelength, 
while the lower column density gas allow most of radiation to escape from the galaxies. 
For example, the SED of halo $(b)$ in Figure~\ref{fig:sed} becomes much redder due to the strong dust extinction, resulting in the large color excess $\EBV=0.37$ for the dust size of $0.05~\rm \mu m$.
This halo however does not satisfy our $\Mab$ selection criteria due to the strong dust extinction, although the intrinsic $\Mab$ is bright with $-21.5$.
On the other hand, halo $(d)$ shows weaker dust extinction than halo $(b)$ despite a very similar halo mass. 
The value of $\EBV$ for $0.05~\rm \mu m$ dust is similar to the observation.
In addition, despite a similar dust extinction at $1700~\A$, halo $(c)$ shows a much smaller $\EBV$ than halo $(d)$, because the SED simply decreases while keeping its shape.
The SEDs can vary due to different viewing angles and inhomogeneous distributions of dust and stars. 
However, the qualitative trends of dust extinction discussed above do not change. 

\begin{figure}
\begin{center}
\includegraphics[scale=0.43]{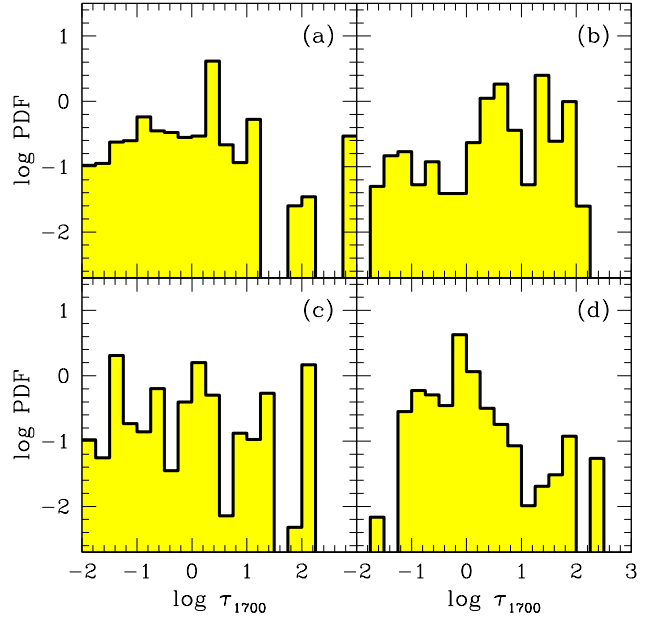}
\caption{
Probability distribution function (PDF) of optical depth at $1700~\A$ for random rays shot from all individual star particles in each halo
 to the grid boundary of halo.  
Panels ($a$) - ($d$) correspond to the same halos in Figure~\ref{fig:sed}. 
The optical depth is estimated for the dust of $\rd=0.05\,\rm \mu m$.
}
\label{fig:stau}
\end{center}
\end{figure}

To investigate the source of variation in $\EBV$ further, we present  the PDF of optical depth at $1700~\A$ from each star particle to the grid boundary for the same halos in Figure~\ref{fig:stau}. 
The large dispersion in optical depth for different stellar positions and viewing angles is caused by the inhomogeneous distribution of stars and dust as shown in Figure~\ref{fig:img}. 
The PDFs of halos $(a)$ and $(b)$ show that a large fraction of line-of-sight have $\tau_{1700} > 1$.
Halo $(c)$ shows a roughly flat distribution, with some at smaller optical depth $\tau_{1700} \lesssim 1$, and some with $\tau_{1700} \sim 100$. 

The higher column densities ($\tau_{1700} \sim 100$) in halos $(b)$ and $(c)$ can cause the dust extinction at $\lambda \gtrsim 10^{4}~\A$ as well, because the optical depth at $\lambda \gtrsim 10^4~\A$ is smaller than $\tauuv$ by two orders of magnitude (Figure~\ref{fig:cross}).  
Moreover, the high column densities absorb most of the flux at UV-to-optical bands, 
whereas the low column densities let the radiation escape without much dust extinction. 
This leads to a decrease in the SED without changing its shape very much and with a smaller $\EBV$ similar to halo $(c)$.
We note that previous semi-analytical or simple models in simulations  used a constant optical depth for all star particles in a galaxy \citep{Dayal09, Kobayashi10}.
Our radiative transfer calculation shows that each galaxy has a large dispersion in the optical depth for individual stellar particles.

\begin{figure}
\begin{center}
\includegraphics[scale=0.4]{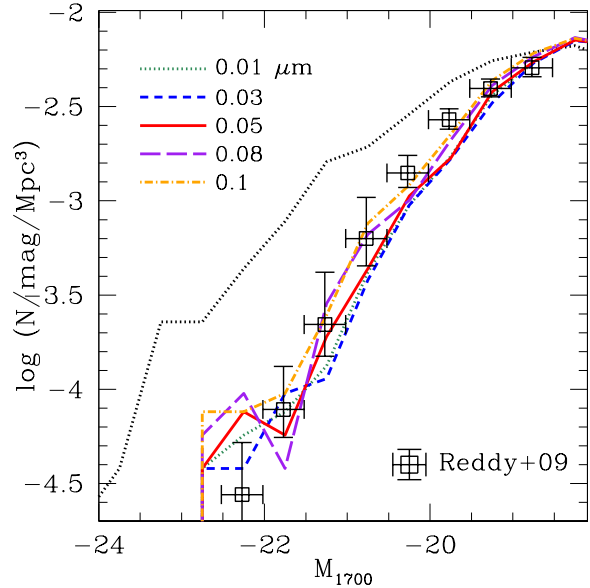}
\caption{
Rest-frame UV luminosity functions of simulated galaxies at $1700~\rm \AA$. 
The black dotted line shows the LFs without dust extinction. 
The colored lines represent the LFs with dust extinction, and the different colors mean different dust sizes.
Open squares indicate the observational data of \citet{Reddy09} at $z\sim 3$. 
}
\label{fig:LF}
\end{center}
\end{figure}

\begin{figure}
\begin{center}
\includegraphics[scale=0.4]{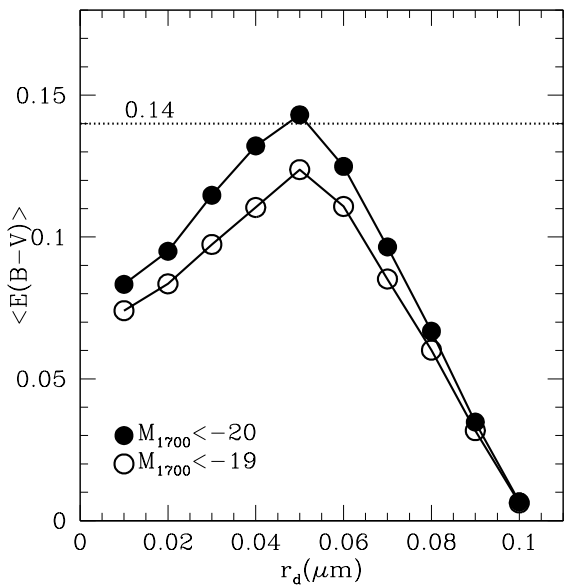}
\caption{
Average $\meanEBV$ of simulated galaxies as a function of dust radius.  
The filled and open circles are the $\meanEBV$ of galaxies with $\Mab \leq -20$ and $\leq -19$, respectively. 
The horizontal dotted line indicates the observed value of $\EBV \sim 0.14$ for LBGs \citep{Shapley01, Reddy08}.  
}
\label{fig:ebv}
\end{center}
\end{figure}


\subsection{Rest-frame UV Luminosity Function}

Having seen some example SEDs and PDFs of $\tau_{1700}$, we then would like to understand the effect of dust extinction on the statistical nature of galaxies. 

Figure~\ref{fig:LF} shows the rest-frame UV LF with and without dust extinction, 
as a function of rest-frame AB magnitude at $1700~\A$ ($M_{1700}$). 
The LF without dust extinction is obviously much brighter than the observational data, as the observations have measured $\EBV \sim 0.14$ \citep{Shapley01, Reddy08}. 
The LFs with dust extinction effect agree with the observational data very well, 
hence our simulations can reproduce the observed LF naturally by combining cosmological simulation with a dust model and the radiative transfer calculation. 
With increasing dust size, the dust extinction decreases somewhat, however
the variation in LF is almost within the error bars of observational data points. 
Therefore the rest-frame UV LF is not so sensitive to the dust size.

We also see in Figure~\ref{fig:LF} that the UV LF converges at the fainter end due to the decreasing dust extinction for fainter galaxies.  This was not the case for the constant extinction model.  
We also point out that the dust content increases with galaxy mass, hence the stellar radiation is efficiently absorbed in higher mass galaxies.   As a result, the dotted line is shifted to the brighter side by a greater amount at the bright-end of the LF from the observed data points. 
Note that the LFs are estimated from a specific viewing angle as the actual observations, 
and the error caused by different viewing angles is negligible for our results. 


\subsection{Dust Extinction}
\label{sec:extinction}

Figure~\ref{fig:ebv} shows the average $\meanEBV$ for galaxies with $\Mab < -20$ and $\Mab < -19$ as a function of dust radius. 
The first selection criteria of $\Mab<-20$ is almost the same as the detection limits of some LBG surveys \citep[e.g.,][]{Ouchi04}. 
In our simulation it roughly corresponds to the halos with $M_{\rm tot} \gtrsim 5 \times 10^{10}~\Msun$,  
which is lower than the typical LBG halo mass of $M_{\rm tot} \sim 10^{12}~\Msun$ \citep{Ouchi04, Adelberger05}.  
Below this limit, the simulated galaxies would have less than $\sim 1000$ SPH particles, 
and the resolution becomes too poor to resolve the halo internal structure of gas and dust distribution. 

Figure~\ref{fig:ebv} shows that $\meanEBV$ has a peak at $\rd=0.05~\micron$ due to the higher ratio of the optical depth between $B$ and $V$-band (Figure~\ref{fig:cross}).
The peak value of $\meanEBV$ is close to the observed one ($\sim 0.14$), and $\meanEBV$ decreases rapidly at smaller or lager dust size. 
Therefore, we suggest that the typical dust size in LBGs at $z \sim 3$ would be $\rd \sim 0.05~\micron$, 
which is similar to that of Type-II supernova dust models \citep{Todini01, Nozawa07}.
Recent simulations showed the radiation properties of LAEs at $z=5.7$ could be reproduced by the $0.05~\micron$ dust  \citep{Dayal09, Dayal11}, implying that the dust in $z=5.7$ galaxies could be explained by the Type-II supernova dust model. 
In this work, we focus on the LBGs at z=3, which would be more massive and evolved than LAEs at $z\sim 6$ typically \citep[e.g.,][]{Gawiser07}.
We will expand our galaxy sample over a wider range of redshift and mass, and investigate the possible redshift evolution of dust properties in our future work. 

\begin{figure}
\begin{center}
\includegraphics[scale=0.4]{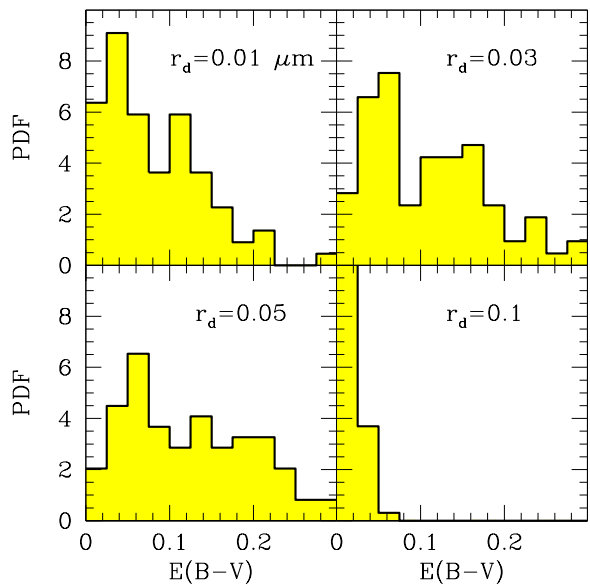}
\caption{
PDF of halos as a function of $\EBV$ for different dust sizes. 
The bin size is 0.025 dex.  
}
\label{fig:pdf}
\end{center}
\end{figure}

The detection limits of recent LBG surveys are becoming somewhat deeper than our fiducial selection criteria with $\Mab =  -19 \sim -18$ \citep{Reddy09}. 
Open circles in Figure~\ref{fig:ebv}  are the results with $\Mab < -19$ limit. 
Since dust extinction of low-mass, faint galaxies is lower due to lower metallicity, 
 $\meanEBV$ with $\Mab < -19$ is somewhat lower than that with $\Mab < -20$. 
Even for $\rd = 0.05~\micron$, it is smaller than the observed value. 
However, other factors could change the values, as we will discuss in Section~\ref{sec:discussion}.

Figure~\ref{fig:pdf} shows the probability distribution function (PDF) of halos as a function of $\EBV$ for different dust sizes.
In the cases of $\rd = 0.01$ and $0.03\,\micron$, the PDFs have peaks at $\EBV \sim 0.05$, and decrease with increasing $\EBV$.
On the other hand, the case of $\rd = 0.05\, \micron$ is less skewed, and distribute over $\EBV \sim 0 - 0.3$.
The broader, more symmetric shape is similar to the observation \citep{Reddy08}, however, our result is somewhat flatter than the observation. 
In the case of $0.1~\micron$ dust, most of galaxies distribute at $\EBV \lesssim 0.05$
due to the small difference of absorption efficiency between $B$ and $V$-band.

\subsection{UV dust extinction vs. Halo Mass}

As shown in the UV LFs, 
UV dust extinction is likely to be stronger for massive halos or brighter ones.
We quantitatively study the relation between UV dust extinction and halo mass.

\begin{figure}
\begin{center}
\includegraphics[scale=0.43]{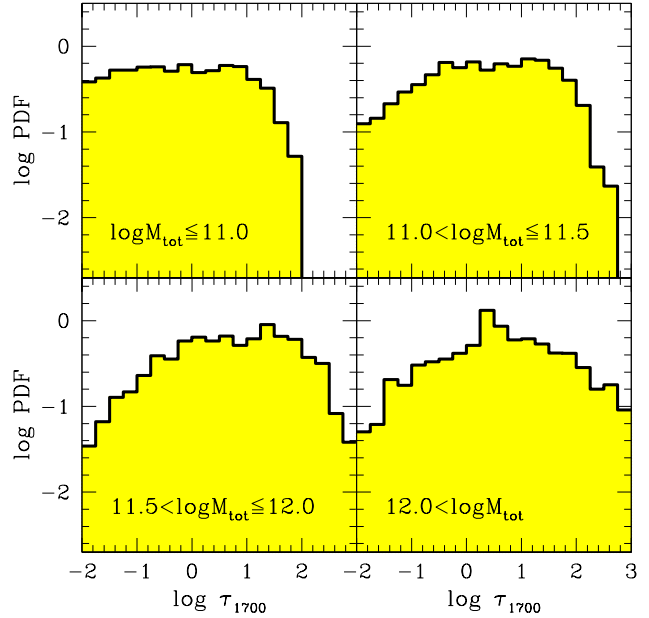}
\caption{
PDF of optical depth at $1700~\A$ for rays shot from all individual star particles to the grid boundary of halo.  
The sample is divided into four halo mass bins, and the optical depths of all star particles in the halos are stacked. 
The optical depth is estimated for the dust of $\rd=0.05\,\rm \mu m$.
}
\label{fig:stau_mass}
\end{center}
\end{figure}

Figure~\ref{fig:stau_mass} shows the PDF  of optical depth at $1700~\A$ for random rays shot from all individual star particles to the grid boundary of a halo.  
We divide our sample into four halo mass bins. 
First, we see the PDF ranges over a wide range of optical depth. 
Even for lower mass galaxies, due to complicated distribution between stars and dusty clouds,
some fraction of star particles have high optical depths of $\tau_{1700} > 10$,
while a large fraction of them have lower values of  $\tau_{1700} \lesssim 1$. 
The PDF clearly shifts to higher optical depth as the halo mass increases. 
For massive halos with $\log\; \Mtot / {\rm \Msun} > 11.5$, the PDF is distributed even at $\tau_{1700} > 100$, while it  decreases at $\tau_{1700} < 0.1$.
This result suggests that a larger fraction of stars are surrounded by more dust and affected by stronger dust extinction as halo mass increases. 

\begin{figure}
\begin{center}
\includegraphics[scale=0.4]{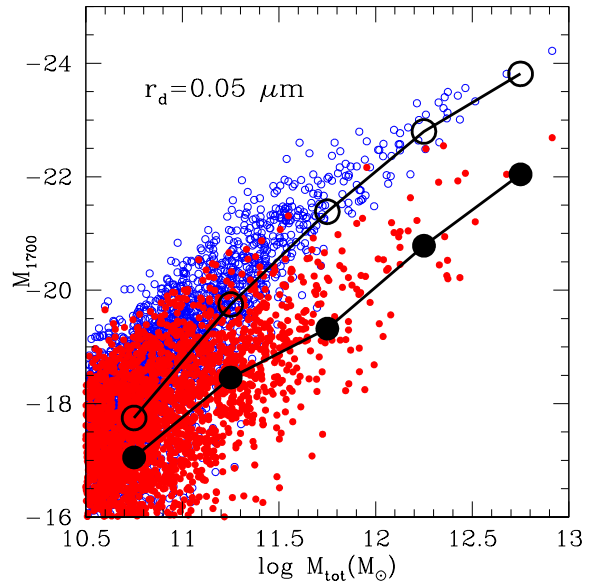}
\caption{
Rest-frame UV magnitude at $1700~\A$ as a function of total halo mass. 
Blue open and red filled circles represent intrinsic fluxes and those after the RT calculation for the dust of $\rd=0.05~\rm \mu m$, respectively. 
Open and filled black circles are the median values of each distribution in each halo mass bin with a bin size of $\Delta \log M=0.5$.
}
\label{fig:mass_M1700}
\end{center}
\end{figure}

In addition, UV fluxes before (blue open circles) and after RT (red filled circles) are shown as a function of halo mass in Figure~\ref{fig:mass_M1700}, for the dust of $\rd=0.05~\rm \mu m$.
Naturally, $\Mab$ increases with halo mass because massive halos have higher SFR at high redshift.  
As we stated above, the dust extinction increases with halo mass due to higher metal content. 
The massive halos with $\Mtot > 10^{12}~\rm \Msun$ suffer from extinction by $\sim 2$ mag,
while low-mass halos with $\Mtot \le 10^{11}~\rm \Msun$ become fainter by only $\sim 0.7$ mag.
Moreover, due to the large dispersion of optical depth shown in Figure~\ref{fig:stau_mass}, 
the UV fluxes have greater dispersion after the RT calculations.

\subsection{$E(B-V)$ vs. Halo Mass, $M_{\rm 1700}$}

Then, we study the dependence of $\meanEBV$ on the halo mass and galaxy luminosity. 

\begin{figure*}
\begin{center}
\includegraphics[scale=0.4]{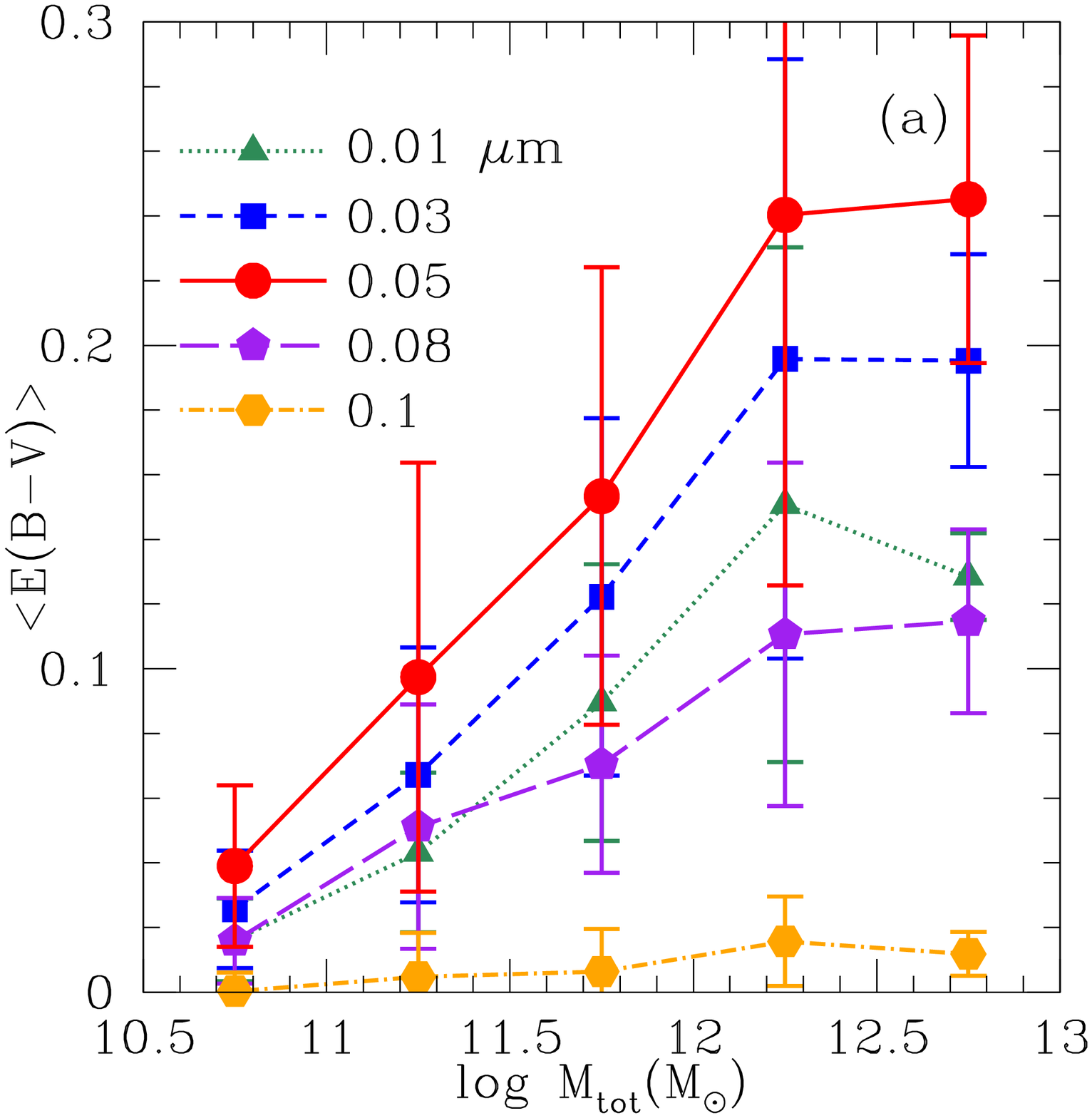}
\includegraphics[scale=0.4]{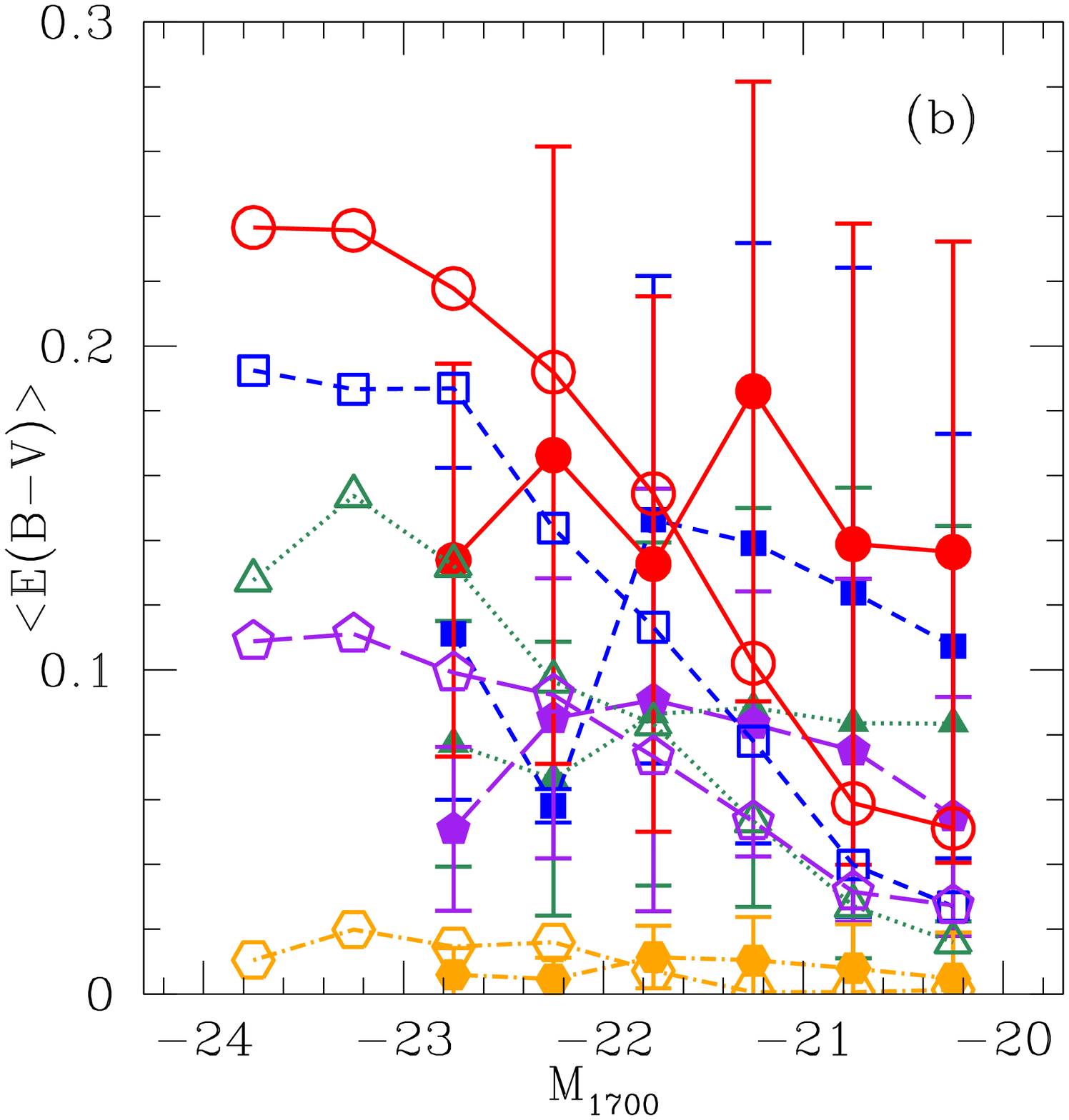}
\caption{
The average $\meanEBV$ of simulated galaxies as a function of halo mass (panel $a$) and $\Mab$ (panel $b$).
The filed symbols are the mean values of each x-axis bin with the bin size of 0.5 dex and $1 \sigma$ error. 
The open symbols are the mean ones to intrinsic $\Mab$ (or without dust extinction), 
on the other hand, the filled symbols are of emergent $\Mab$ (with dust extinction).
}
\label{fig:ebv_mass}
\end{center}
\end{figure*}

In Figure~\ref{fig:ebv_mass}(a), we plot $E(B-V)$ vs. total halo mass for those with $M_{\rm 1700}<-20$ after the RT calculation, and compute the average $\meanEBV$ in each bin. 
The result shows that $\meanEBV$ increases with the halo mass due to higher dust mass in more massive halos. 
In each bin of halo mass, there is significant dispersion in $\meanEBV$ due to different distribution of metals and dust among halos. 
As expected from the results of Fig.~\ref{fig:cross}, the case of $0.05\,\micron$ dust have the highest $\meanEBV$, reaching $\sim 0.25$ at $\Mtot > 10^{12}~\Msun$.
As seen in Equation~\ref{eq:ebv}, the $E(B-V)$ becomes zero if there is no dust extinction or the dust extinction curve is flat at around $4400~\A \lesssim \lambda \lesssim 5500~\A$.
In the case of $0.1\,\micron$ dust, the value of $\meanEBV$ is very small due to the almost flat extinction curve, and does not exceed $\sim 0.02$. 

Figure~\ref{fig:ebv_mass}(b) shows the $\meanEBV$ as a function of $\Mab$: the open triangles are plotted with the intrinsic $\Mab$ values without dust extinction effect (hereafter $M_{\rm 1700, int}$), and the filled circles are plotted with $\Mab$ after considering the dust extinction effect (hereafter $M_{\rm 1700, after}$).
In our simulations, the intrinsic luminosity and SFR is roughly proportional to the total halo mass and galaxy mass, therefore the values of $\meanEBV$ are higher for more massive, brighter halos as shown in Figure~\ref{fig:ebv_mass}(b) (open triangles). 
However, after the dust extinction effect is considered, the massive bright galaxies move to the fainter bins, 
therefore the fainter $\Mab$ bins contain both faint low-mass galaxies with little dust, as well as the massive bright galaxies with significant dust. 
This effect leads to almost constant $\meanEBV$ when plotted as a function of $M_{\rm 1700, after}$. 
Therefore, $\meanEBV$ at fainter $M_{\rm 1700, after}$ can be high and have large dispersion in Figure~\ref{fig:ebv_mass}(b) (filled circles). 
The weak dependence of $\EBV$ to $M_{\rm 1700, after}$ is similar to the observed data \citep{Reddy08}.
This trend was also reported by the previous theoretical work with the semi-analytical model of galaxy formation and the simple RT model \citep{Gonzalez-Perez13}. 

\begin{figure*}
\begin{center}
\includegraphics[scale=0.4]{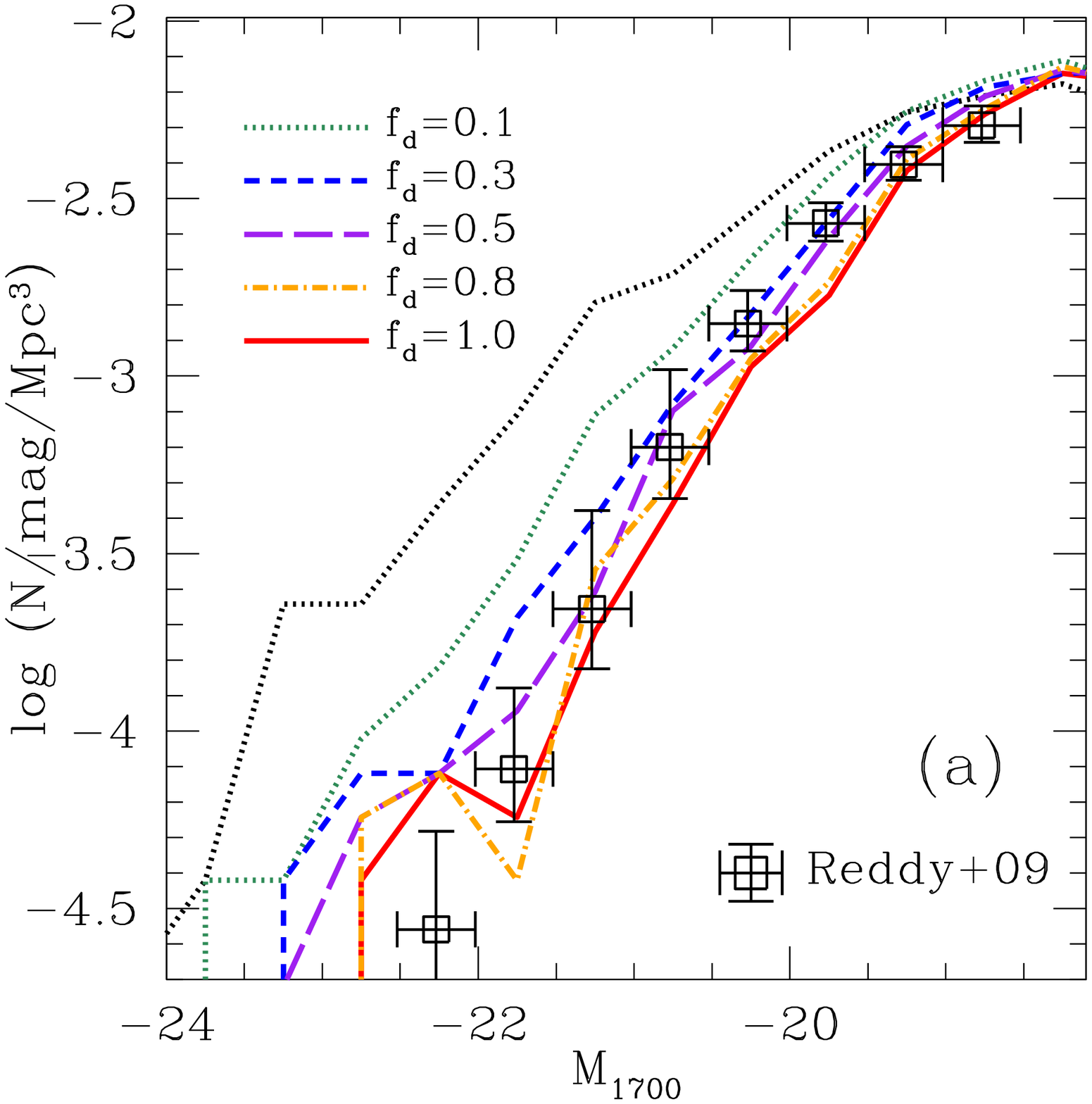}
\includegraphics[scale=0.4]{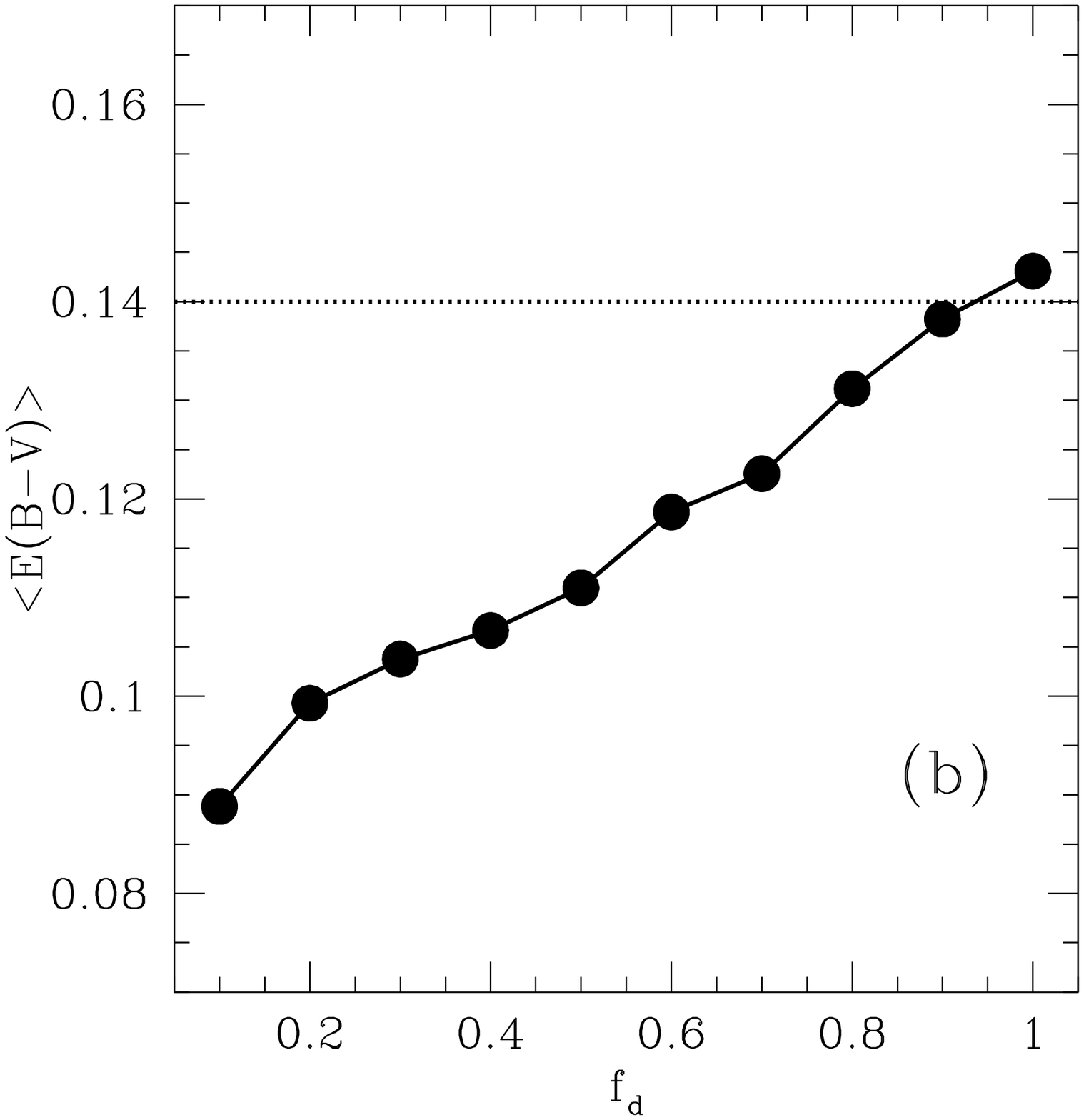}
\caption{
{\it Panel (a)}: Rest-frame UV LFs of simulated galaxies with different normalization parameter $\fdd$ for dust-to-metal mass ratio (see Eq.~\ref{eq:dusttogas}). 
{\it Panel (b)}: The mean $\EBV$ of simulated galaxies with $M_{\rm 1700} < -20$ as a function of $\fdd$. The dust size of $\rd = 0.05\,\micron$ is assumed here.
} 
\label{fig:ebv_Dfactor}
\end{center}
\end{figure*}

%
%

\section{Discussion}
\label{sec:discussion}

\subsection{Dust-to-metal mass ratio}

\begin{figure*}
\begin{center}
\includegraphics[scale=0.4]{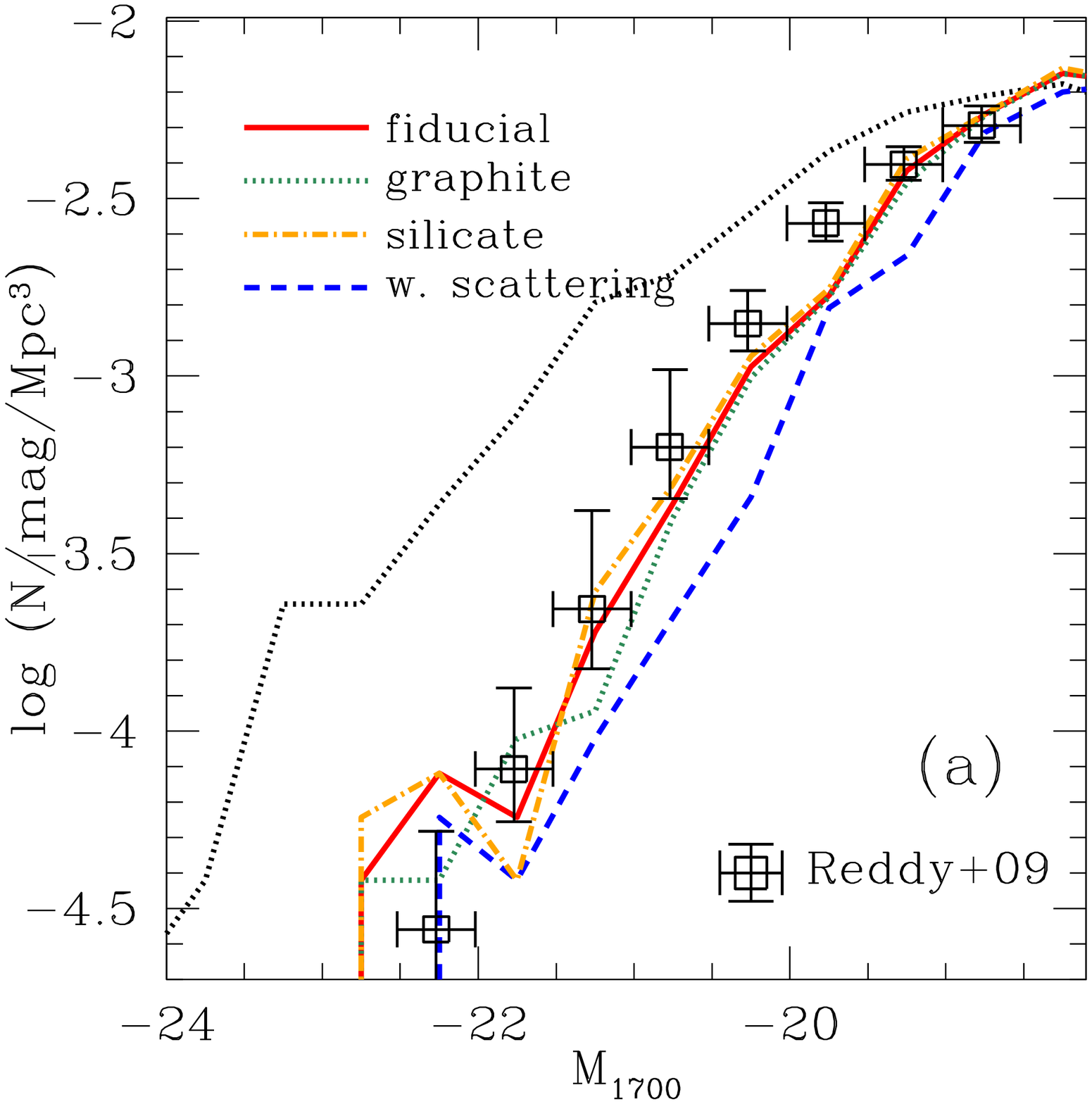}
\includegraphics[scale=0.4]{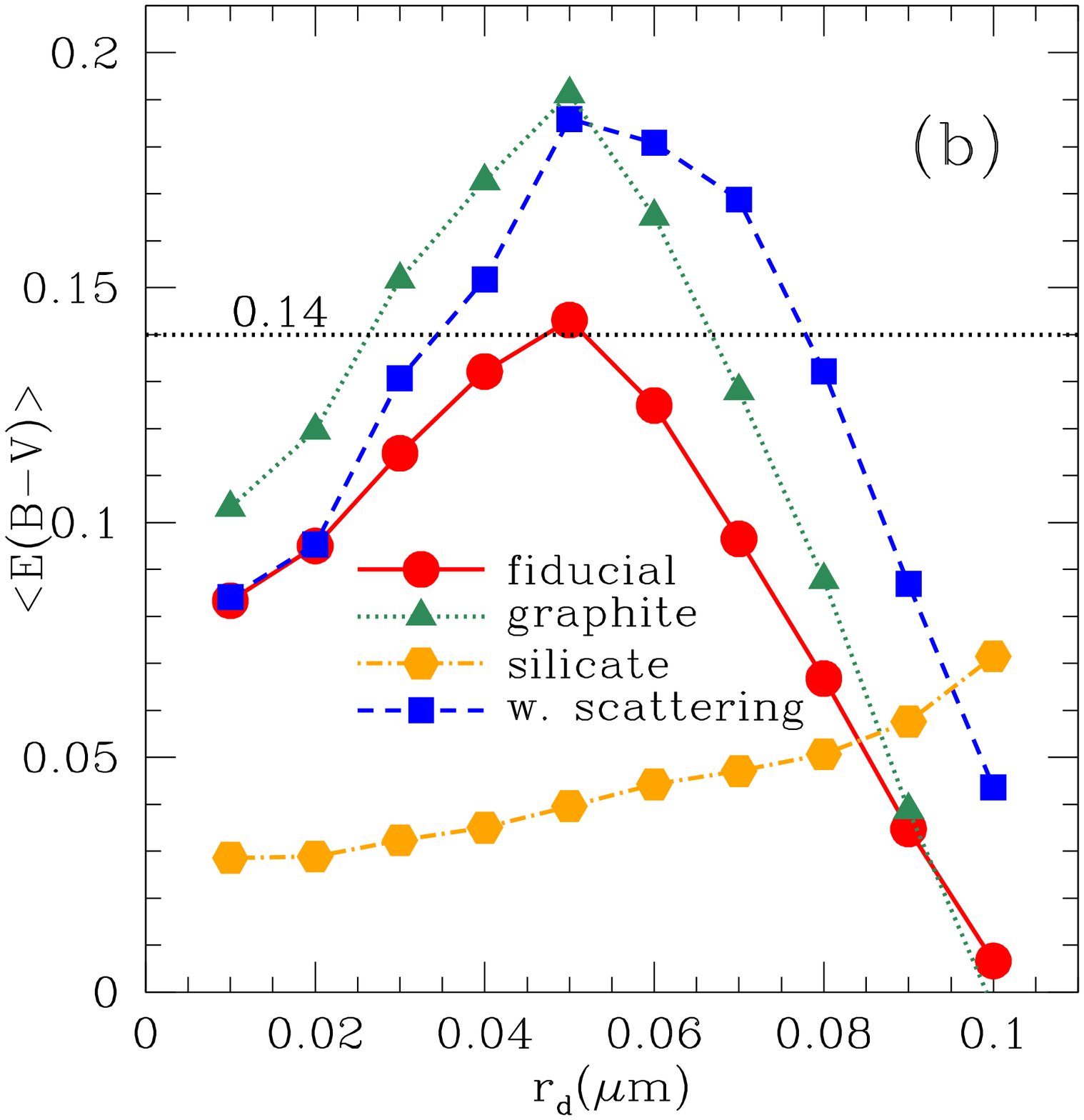}
\caption{
{\it Panel (a): } Rest-frame UV LFs of simulated galaxies with different dust compositions. 
The blue line also shows the LF with maximum scattering effect. 
Here the $0.05\,\micron$ dust is used. 
{\it Panel (b): } Mean $\EBV$ as a function of dust size with different compositions.  The blue line shows the maximum scattering case. 
}
\label{fig:ebv_comp}
\end{center}
\end{figure*}

As we discussed in Equation~(\ref{eq:dustgas}), 
we have taken the  dust-to-gas ratio of $\kappa = 0.01$ at solar abundance so far, 
which is similar to the value for Milky Way and other local galaxies \citep{Draine07}.
However, at higher redshift, a large fraction of dust can be destroyed due to the sputtering process in supernova shock waves \citep{Nozawa07}.  
In addition, the contribution from AGB stars is little at high redshift. 
Therefore, the value of $\kappa$ could be lower at higher redshifts than the local value. 

Here we examine the effect of varying $\kappa$ with a modulation factor $\fdd$ defined by
\begin{equation}
M_{\rm dust} = \fdd \times 0.01 M_{\rm gas} \left (\frac{Z}{\Zsun} \right ). 
\label{eq:dusttogas}
\end{equation}
As shown in Equation~(\ref{eq:dustgas}), our fiducial model assumes that the dust-to-gas mass ratio increases 
with metallicity based on the constant dust-to-metal mass ratio, $M_{\rm dust} / M_{\rm metal} = 0.5$.
The above $\fdd$ controls the dust-to-metal mass ratio, for example, $\fdd=0.5$ means the dust-to-metal ratio is reduced to 0.25.

Figure~\ref{fig:ebv_Dfactor}(a) shows the UV LFs computed with different $\fdd$. 
The LF shifts to the brighter side slightly with decreasing $\fdd$ due to the lower amount of dust. 
However, the LFs with $\fdd=0.3 - 1.0$ are mostly within the error bars of the observed data, and the comparison of LFs cannot constrain the values of $\fdd$.  
Our comparison only suggests $\fdd > 0.1$.

On the other hand, $\meanEBV$ depends on $\fdd$ sensitively, and linearly increases with it as shown in Figure~\ref{fig:ebv_Dfactor}(b).  
Our result suggests that $\fdd \gtrsim 0.8$ is required, in order to reproduce the observed value of $\EBV \sim 0.14$. 
Therefore,  even at $z \sim 3$, we suggest that the relation of dust-to-metal mass ratio as a function of metallicity is close to the local relation. 
Recent observation using GRB aftergrows also suggested higher dust-to-metal mass ratios in high-redshift galaxies \citep{Zafar13}, which is consistent with our simulation results. 


\subsection{Effect of Dust Composition}

So far we have assumed the dust composition of graphite and silicate with a 1:1 mass ratio (fiducial model). 
In this section, we discuss the impact of different mass ratios. 
As a test, we use simple models that all dust are either graphite or silicate. 
Figure~\ref{fig:ebv_comp}(a) shows that the rest-frame UV LFs with only silicate or graphite dust 
do not differ so much from the 1:1 mix fiducial model, and they all agree relatively well with the observed data.

\begin{figure}
\begin{center}
\includegraphics[scale=0.4]{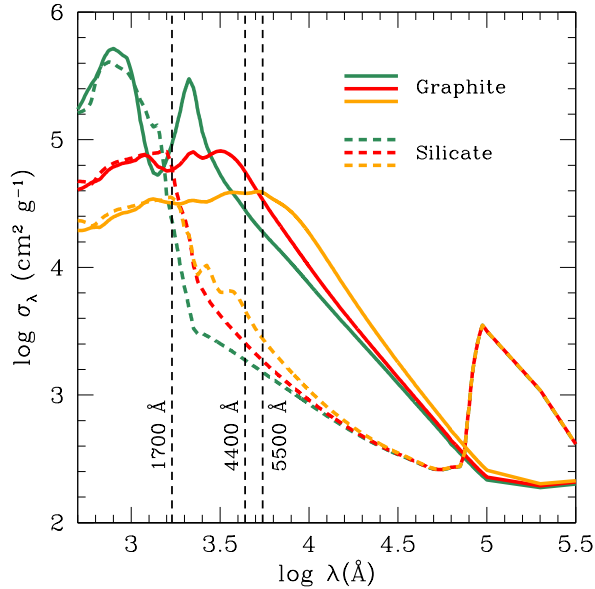}
\caption{
 Absorption efficiency per unit dust mass $\sigma_\lambda$ as a function of wavelength. 
Different colors represent different dust sizes: $0.01~\rm \mu m$ (green), $0.05~\rm \mu m$ (red), and $0.1~\rm \mu m$ (orange), respectively.
Solid and dash lines show dust models which consists of either graphite or silicate alone, respectively. 
}
\label{fig:sigma_silicate}
\end{center}
\end{figure}

On the other hand, $\meanEBV$ is affected by the dust component significantly, as shown in Figure~\ref{fig:ebv_comp}(b).
The $\meanEBV$ with graphite dust (green line) is higher than the fiducial model, 
while with silicate dust it is very small and almost constant.  
One might wonder why the silicate-only model can reproduce the UV LF so well with such little $\meanEBV$. 
Silicate dust has small $\Qabs$ at optical wavelengths \citep{Draine84, Laor93}. 
Figure~\ref{fig:sigma_silicate} shows the absorption efficiency of dust which consists of either graphite or silicate alone. 
For example, $0.05~\micron$ silicate dust has $\Qabs \sim 0.04$ at $5000~\A$, while it is $\sim 1.2$ at $1700~\A$.
Since $\EBV$ is proportional to $\log [\exp(-\tau_{\rm V})/ \exp(-\tau_{\rm B})]$, where $\tau_{\rm B}$ and $\tau_{\rm V}$ are the optical depth at $B$- and $V$-band,
the $\EBV$ becomes very small when $\tau_{\rm V}, \tau_{\rm B} \ll 1$.
On the other hand, the $\Qabs$ of silicate at $1700~\A$ is similar to that of graphite, hence it shows similar UV LFs. 
As a result, while silicate dust can reproduce the observed UV LF with a moderate dust extinction at $1700~\A$, it fails to reproduce the observed $\EBV$.
Therefore, we conclude that the graphite dust is mainly responsible for causing extinction, and it can increase the extinction up to $\meanEBV \sim 0.19$ with $\rd=0.05\,\micron$ by itself. 
Our result suggests that, in order to reproduce the observed level of $\EBV$, 
at least more than half of dust should be composed of graphite. 


\subsection{Effect of Scattering}
\label{sec:scattering}

Finally we discuss the impact of scattering. 
So far we have considered only the dust absorption, however 
in reality, some fraction of UV continuum photons can be scattered by dust, and then absorbed. 
In this section, we study the maximum effect of scattering by assuming that all scattered photons are absorbed, or equivalently that they are scattered out of our line-of-sight and no additional scattered photons enter our viewing angle. 
The fiducial mixed dust model with silicate and graphite is used for the study of scattering in this section. 

The blue line in Figure~\ref{fig:ebv_comp}(a) shows the rest-frame UV LF with the maximum scattering effect for $0.05\,\micron$ dust. 
The LF with maximum scattering is much fainter than the observed data due to the strong extinction, and it is clearly inconsistent with Reddy's data. 

The blue line in Figure~\ref{fig:ebv_comp}(b) shows the $\meanEBV$ as a function of dust size with the maximum scattering effect. 
It peaks at $\rd=0.05\,\micron$ similarly to the one without scattering (Fig.~\ref{fig:ebv}), but the peak value is higher with $\meanEBV \sim 0.19$. 
This suggests that it is difficult to reproduce the observed UV LF and $\EBV$ simultaneously with the maximum scattering effect. 
One possible way to fit the observations even with the maximum scattering effect 
would be reducing $\kappa$, as it would decrease the dust extinction at $1700~\rm \AA$ and $\EBV$ as we saw in Fig.~\ref{fig:ebv_Dfactor}(b).


%
%

\section{SUMMARY}
\label{sec:summary}

In this paper, we investigate the dust properties in the LBGs at $z=3$ by combining cosmological simulations with multi-wavelength RT calculations. 
We compute the RT from all star particles in 2800 massive halos at $z=3$ that contain LBGs, and compared the rest-frame UV LF and $E(B-V)$ with observations. 
We only examine galaxies brighter than $\Mab = -20$ or $-19$, 
corresponding to the detection limits of recent LBGs surveys. 

The most significant result is that our simulation can naturally reproduce the observed rest-frame UV LF with a dust model that work on an individual particle level. 
Earlier works using cosmological simulations have either adopted a constant extinction value \citep[e.g.,][]{Nagamine04e, Night06} or an ad hoc power-law relation between galaxy metallicity and the extinction with some scatter \citep[e.g.,][]{Finlator06}. 
Our present work is an important step forward from those previous treatments, in the sense that we now compute the extinction of each galaxy based on the dust properties of each gas particle, utilizing the RT  technique. 
A similar approach using RT was taken by the SUNRISE code \citep{Jonsson06} and the $\rm ART^{2}$ code \citep{Li08, Yajima12b, Yajima12f}.
They computed the SEDs  and surface brightness of galaxies with a specific dust model, e.g., the Milky Way \citep{Weingartner01} or supernova \citep{Todini01}.
In the present work, considering the large uncertainties of dust properties in high-redshift galaxies, we tested many dust models with varying parameters using simple ray-tracing method, and calculated the SEDs of numerous galaxies. 

Another point is that the SUNRISE and $\rm ART^{2}$ take the scattering process into account based on the albedo of dust using random numbers. 
On the other hand, we considered only an upper limit case where all scattered photons are absorbed by dust and disappears from the viewing angle (\S~\ref{sec:scattering}). 
Overall, our calculation method is simpler than the SUNRISE or ART$^2$ 
methods which allows us to probe a wider range of dust parameters.

We also examined the sensitivity of $\EBV$ to the dust properties, namely their composition and grain size. 
Based on the comparison between our simulations and observations, 
we suggest that the dust in LBGs at $z=3$ likely have the following properties: 
\begin{itemize}
\item Typical dust size is $\rd \sim 0.05\,\micron$, which is similar to that originated from Type-II supernova.
\item Dust-to-metal mass ratio is close to that of local galaxies.  
\item More than half of dust consists of graphite, rather than silicates. 
\end{itemize}

In the present work, our fiducial dust model adopts the following assumptions: a spherical shape,  the relationship of $M_{\rm dust}= 0.01 M_{\rm gas} (Z/ \Zsun)$, and the composition of graphite and silicate with a 1:1 mass-ratio.  
Our best-fit model is $\rd \sim 0.05\,\micron$ and $\kappa=0.01$, which is close to the local value. 
However, a different mass ratio of composition and scattering may change the best-fit model to somewhat lower dust-to-metal mass ratios
and different dust sizes (other than $\rd = 0.05\,\micron$).
Future multi-wavelength observations and theoretical simulations will be able to give us more information to constrain these details further.  
In the future, we will extend our work to a wider redshift range, and investigate the redshift evolution of dust properties. 

The mixing and distribution of dust in galaxies is still a difficult and unresolved issue in current theoretical models of galaxy formation.
In particular, the lack of mixing in SPH codes relative to mesh codes has been noted before by several authors \citep[e.g.,][]{Agertz07}, and new SPH formulations to remedy this problem have been developed recently \citep[e.g.,][]{Saitoh13, Hopkins13}.  
Furthermore, \citet{Vogelsberger12} compared the gas distribution in galaxies between SPH and the moving mesh code, {\sc arepo}, and showed that 
the gas distribution in the SPH run was more clumpy than in {\sc arepo}.
Such clumpy gas structure may make the dispersion of $\EBV$ greater, because some fraction of viewing angles can interact with high-density gas clumps, resulting in strong dust attenuation.
\citet{Hopkins13} also showed that, with the newer formulation of SPH, the clumpy structure of previous SPH simulations disappears due to more efficient mixing of gas.  
On the other hand, there are many simulation results from AMR codes that show clumpy gas distributions in high-redshift galaxies \citep[e.g.,][]{Ceverino13}.  
We plan to repeat our calculations with the new version of SPH in the near future, and examine how the clumpy gas distribution may have affected our results on the $\EBV$ distribution. 
Future observations will also resolve the gas structure in higher redshift galaxies better, and we will be able to make more robust comparisons and develop a better model of galaxy formation. 
We note in passing however, that we are currently limited more by the speed of the RT code, which prohibits us to use large grid sizes for massive halos, even when we have higher hydrodynamic resolution in SPH simulations.  

Any results on dust attenuation would be sensitive to dust distribution, and the details of the feedback prescription will also be important for dust distribution. 
Our current work assumed that dust distribution coupled metal, and our simulation used the hybrid Multi-component Variable Velocity (MVV) wind model \citep[][see Section~\ref{sec:model}]{Choi11} to transport metals from Type-II supernovae into intergalactic medium via galactic outflows.  
The MVV wind model was more successful in reproducing the observed IGM metallicity compared to the earlier constant speed wind model, but there is still room for further improvement in our feedback model, such as the inclusion of early stellar feedback by stellar winds and radiation pressure, which we plan to investigate in the future.

%
%
\section*{Acknowledgments}
We are grateful to Y. Li  for valuable discussion and comments, 
and to Volker Springel for providing us with the original version of GADGET-3.
This work is supported in part by the President's Infrastructure Award from UNLV, 
and the HST grant AR-12143.01-A, provided by NASA through a grant from the Space Telescope Science Institute, which is operated by the Association of Universities for Research in Astronomy, Incorporated, under NASA contract NAS5-26555.
This research is also supported by the NSF through the TeraGrid resources 
provided by the Texas Advanced Computing Center (TACC).

%
%




\label{lastpage}

\end{document}